\documentclass[aip,apl,twocolumn,superscriptaddress,groupedaddress]{revtex4}  % for review and submission
\usepackage{graphicx}  % needed for figures
\usepackage{dcolumn}   % needed for some tables
\usepackage{bm}        % for math
\usepackage{amssymb}
\usepackage{amsmath}\allowdisplaybreaks
\usepackage{epstopdf}
\usepackage{color}
\setcitestyle{super}
\usepackage{enumerate}
\usepackage{mathrsfs}
\usepackage{hyperref}
\begin{document}

\title{Relationship of Time Reversal Symmetry Breaking with Optical Kerr Rotation}

\author{Alexander D. Fried}
\affiliation{Department of Physics, Stanford University, Stanford, CA 94305}
\date{\today}
\begin{abstract}
We prove an instance of the Reciprocity Theorem that demonstrates that Kerr rotation, also known as the magneto-optical Kerr effect, may only arise in materials that break microscopic time reversal symmetry.  This argument applies in the linear response regime, and only fails for nonlinear effects.  Recent measurements with a modified Sagnac Interferometer have found finite Kerr rotation in a variety of superconductors.  The Sagnac Interferometer is a probe for nonreciprocity, so it must be that time reversal symmetry is broken in these materials.
\end{abstract}
\maketitle

\section{Introduction}

Recently, there has been controversy\cite{Pavan2,Mineev, Mineev3,Pavan,Arfi,Yakovenko1,Yakovenko2,Varma} surrounding the apparent measurement of finite polarization rotation, also known as the magneto-optical polar Kerr Effect, in optical reflection measurements off of a variety of High Tc superconductors.\cite{hovo0,hovo}  The polar Kerr effect is characterized by the Kerr angle, $\theta_K$, which is the difference in phase angle delays yielded by oppositely circularly polarized plane-wave beams of light upon normal incidence reflection from a sample.  While the Kerr effect is often associated with magnetic materials,\cite{argyres,Orenstein,Varma} it has been suggested that the observations of Karapetyan \emph{et al}\cite{hovo0,hovo} are more consistent with cholesteric order\cite{Pavan}\textemdash an order characterized by mirror asymmetry about any plane.  Although these measurements convincingly demonstrate a real and novel effect, we argue that the interpretation of cholesteric order is flawed.  This discussion has broached a much more long-standing controversy\cite{Svirko1,Svirko2,Svirko3,Svirko4,Svirko5,Svirko6,Svirko7,Svirko8,Ginzberg0,Agranovich,shelankov,dodge,armitage,Gridnev,Canright,Canright3,Cardona2,silverman,silverman2,silverman_expt,Halperin,Agronovich2,Schlaghneck,Vinogradov,Gridnev2,bokut,Fedorov,Nelson1,Nelson2,Cardona,Kamenetskii}
 regarding the correct form of gyrotropic electromagnetic constitutive relations, and whether Kerr rotation is allowed by general optically active media\textemdash media that break only mirror symmetry about a plane containing the surface normal.\cite{LandauEM}  In this paper, we prove a general theorem that guarantees that the observation of Kerr rotation must always imply that it is microscopic time reversal symmetry that is broken.  

The idea that the Kerr effect implies microscopic time reversal symmetry breaking has been argued by a number of authors,\cite{shelankov,dodge,armitage,Canright,Canright3,Cardona,Cardona2,silverman,silverman2,silverman_expt} but those conclusions made by Halperin\cite{Halperin} provide a useful introduction.  He considers a plane-wave source and adjacent detector, both at fixed distance along the z-axis to the sample at $z=0$.  The distance is great enough such that the source and detector may as well be considered on top of each other.  Let $R_{++}$ and $R_{--}$ be the reflection amplitudes for circular polarization states reflecting into circular polarization states with the same sense of rotation; for incident and reflected rays propagating along the $\hat{\mathbf{z}}$ axis, $\pm$ refers to the polarization state of the electric field given by $\mathbf{E}_\pm=\Re \{ \tfrac{1}{\sqrt{2}}(\hat{\mathbf{x}}\pm i\hat{\mathbf{y}})e^{i k_z z-i\omega t}\}$ as $z\rightarrow \infty$.  $\pm$ also may be understood as the sign of the spin angular momentum of the light with respect to the $\hat{\mathbf{z}}$ axis and independent of the direction of propagation, $k_z\hat{\mathbf{z}}$; $\pm$ is \emph{not} the helicity.  By application of Onsager's relations, Halperin demonstrates that if the material is time reversal symmetric, then $R_{++}=R_{--}$.  Since the Kerr angle is $\theta_K=\tfrac{1}{2} (\arg R_{++}-\arg R_{--})$, it will be zero when the material preserves time reversal symmetry.  While his argument is satisfactory, it is deserving of a more rigorous discussion.

We begin as Halperin does.  Consider a general measurement of the reflection amplitudes where the pair of sources, each collocated with a detector, are positioned arbitrarily with respect to a sample and each other.  Let the sources be of arbitrary shape, but emit light, which, in the absence of all other sources or scatterers, appears as a circularly polarized plane wave at infinite distance.  In the presence of scatterers, the emitted field may still be described as having a circular polarization state $\pm$ near the source, if not as a plane wave.  We consider the experiment where light of the $+$ polarization state is emitted at a source located at $\mathbf{r}_1$ and the $+$ component of the reflected wave is measured at a detector at $\mathbf{r}_2$.  Let, also, light of the $-$ polarization state, be sourced at $\mathbf{r}_2$ and the $-$ component of the polarization be measured at $\mathbf{r}_1$.  This is accomplished if the collocated detectors are such that they signal the arrival of a photon in the time reverse of the quantum optical state initially formed at the respective source.  Again, the Kerr angle is the measured difference in complex arguments of the two propagation amplitudes.  In the limit of $\mathbf{r}_1\rightarrow\mathbf{r}_2\rightarrow\infty$, the measured reflection amplitudes are the same as $R_{++}$ and $R_{--}$ described by Halperin.

We will demonstrate that, for the measurement described above, when the instrumentation and the sample consist of materials that are all time reverse symmetric, the electromagnetic propagation amplitude from $\mathbf{r}_1$ to $\mathbf{r}_2$ will always be identical to that for propagation from $\mathbf{r}_2$ to $\mathbf{r}_1$.  It then follows that the Kerr angle will also be zero when there is time reversal symmetry, and that broken mirror symmetry, alone, can not give rise to Kerr rotation.

\section{Propagators for Optical Measurements}

Photon Green's functions describe optical measurements.   In the macroscopic limit, the light emitted from a source and measured by a detector is modeled by the retarded Green's function for the macroscopic Maxwell's equations:
\begin{align*}
&\nabla\times\mathbf{E}=-\tfrac{1}{c}\partial_t\mathbf{B} &&\nabla\cdot\mathbf{D}=\rho_f\\
&\nabla\times\mathbf{H}=\tfrac{1}{c}\partial_t\mathbf{D}+\tfrac{4\pi}{c}\mathbf{J} &&\nabla\cdot\mathbf{B}=0
%\mathbf{E}+\int \tilde{\chi}(t,\mathbf{r},t',\mathbf{r}')\mathbf{E}(t',\mathbf{r'})d\mathbf{r}'dt'\\
\end{align*}
Where $\mathbf{B}=\nabla\times \mathbf{A}$ and $\mathbf{E}=-\frac{1}{c}\partial_t \mathbf{A}$ in radiation gauge.  At optical frequencies, it is sufficient to describe the material's response with just a dielectric susceptibility tensor, $\tilde{\chi}(t_2,\mathbf{r}_2,t_1,\mathbf{r}_1)$.\cite{pershan,LandauEM}  The retarded Green's function, $\tilde{G}^\text{ret}$, relates the source current, $\mathbf{J}=(\text{J}_x,\text{J}_y,\text{J}_z)$, to the macroscopic vector potential, $\mathbf{A}$:
\begin{equation}
\mathbf{A}(t_2,\mathbf{r}_2)=\frac{4\pi}{c}\int\tilde{G}^\text{ret}(t_2,\mathbf{r}_2,t_1,\mathbf{r}_1)\mathbf{J}(t_1,\mathbf{r}_1)dt_1 d\mathbf{r}_1
\label{responsefunc}
\end{equation}

Precise statements of the symmetries of the electromagnetic field and its measurement entail that the reflection amplitudes be considered quantum mechanically.\cite{Glauber,Fetter}    The quantum electrodynamic field measured at $(t_2,\mathbf{r}_2)$ by a point dipole detector, aligned to the $\mu$ linear polarization state, will be $\hat A_\mu(t_2,\mathbf{r}_2)\vert 0\rangle$, where $\hat{A}_\mu(0,\mathbf{r})=\hat{A}^\dagger_\mu(0,\mathbf{r})$ is the position-space field operator and $\vert 0\rangle$ is the vacuum state.  Likewise, supposing a point-like dipole source creates a $\nu$ linearly polarized photon at $(t_1,\mathbf{r}_1)$, then the quantum field it initially forms will be $\hat A_\nu(t_1,\mathbf{r}_1)\vert 0\rangle$.  For $t_2>t_1$, the amplitude for free-space propagation between the source and receiver is given by:
\begin{equation*}
\langle 0\vert\hat A_\mu(t_2,\mathbf{r}_2)\hat A_\nu(t_1,\mathbf{r}_1)\vert 0\rangle=\delta_{\mu\nu}\frac{ \delta\big( t_2-t_1-\tfrac{1}{c}\vert \mathbf{r}_2-\mathbf{r}_1\vert\big)}{4\pi\vert\mathbf{r}_1-\mathbf{r}_2\vert}
\end{equation*}
Squared, this is the transition probability density for the detection of a photon at time $t_2$ given its creation at $t_1$.\cite{Glauber}  When the sources are of a single frequency $\omega$, the phase delay, as used to define the Kerr angle, is the complex argument of the propagator in the frequency-position domain: $\tilde G^\text{ret}(\omega;\mathbf{r}_2,\mathbf{r}_1)=\int \tilde G^\text{ret}(t_2,\mathbf{r}_2,t_1,\mathbf{r}_1)e^{i\omega (t_2-t_1)} dt_2 dt_1$

When the light is interacting with matter, then to lowest order, the linear response of the macroscopic field at the detector, $\mathbf{A}(t,\mathbf{r}_2)$, for $\mathbf{r}_2$ outside of the material, to an optical source at $\mathbf{r}_1$, also outside of the material, is given by Equation \ref{responsefunc}.\cite{Fetter}  The retarded Green's function is obtained by complex conjugating the negative frequency part of the following time-ordered propagator:
\begin{equation}
G^\text{F}_{\mu\nu}(t_2,\mathbf{r}_2,t_1,\mathbf{r}_1)=\langle g\vert \text{T}\big[\hat{A}_\mu (t_2,\mathbf{r}_2) \hat{A}_\nu(t_1,\mathbf{r}_1)\big]\vert g\rangle
\label{green}
\end{equation} where $\text{T}$ is the time-ordering operator for photons: $\text{T}\big[\hat{A}_\mu (t_2,\mathbf{r}_2) \hat{A}_\nu(t_1,\mathbf{r}_1)\big]=\theta(t_{2}-t_{1})\hat{A}_{\mu}(t_{2},\mathbf{r}_{2})\hat{A}_{\nu}(t_{1},\mathbf{r}_{1})+\theta(t_{1}-t_{2})\hat{A}_{\nu}(t_{1},\mathbf{r}_{1})\hat{A}_{\mu}(t_{2},\mathbf{r}_{2})
 $.   We choose to focus on the time-ordered propagator just to emphasize how propagators are calculated from quantum perturbative methods.

The expectation value is taken with respect to the many-body ground state, $\vert g\rangle=\underset{t\rightarrow-\infty}{\text{lim}}\hat g^\dagger(t)\vert 0\rangle$ of the whole system.  This ground state includes the material, the environment and any instrumentation.  If the system is at finite temperature, then a Boltzmann weighted sum of propagators, evaluated with respect to the stationary states of the system is used in lieu of the above.  In this way, even incoherent optical sources\cite{shelankov} may be described.

In assuming that the measurement is described exactly by Equation \ref{green}, it is implied that the source is the perturbation to the full Hamiltonian of the world, $\hat{H}$, which describes the light, the material and the detectors.  The perturbing source, $\mathbf{J}(t,\mathbf{r})$, is slowly turned on from zero at $t=-\infty$ and slowly turned off at $t=\infty$.  It is also assumed that the sample, the source and receiver do not interact in any way other than by the scattered light; this is tantamount to requiring that the operators $\hat A_\mu(t,\mathbf{r}_2)$ and $\hat A_\nu(t,\mathbf{r}_1)$ commute with each other, and with $\hat g(t)$ and $\hat g^\dagger(t)$ at equal times.  These are the same conditions requisite for application of the Kubo Formula, and results similar to those in the next section appear in many texts in connection with it.\cite{Martin}

\section{The Reciprocity Theorem}
\label{ProofTheorem}

We will prove that, if time reversal symmetry is respected, then no Kerr rotation is observed, by showing that this symmetry condition implies that the propagator for $+$ polarized light traversing from $\mathbf{r}_1\rightarrow\mathbf{r}_2$ and the propagator for $-$ polarized light traversing from $\mathbf{r}_2\rightarrow\mathbf{r}_1$ are identical.  Of central importance is that the measurement is performed with collocated sources and detectors, which create or destroy photons in states that are the time reverse of each other.  This condition is clearly true for the two point-like dipole sources/detectors, located at $\mathbf{r}_1$ and $\mathbf{r}_2$, considered in this discussion.  We later describe an example of how this is achieved in practice.

The anti-linear time reversal operator,\cite{Weinberg,Sachs,Coester,Schwinger} $\mathcal{T}$, commutes with the Hamiltonian, $\hat H$; $\mathcal{T} \hat{H}\mathcal{T}^\dagger=\hat{H}$, but still inverts the time-evolution operator, $\mathcal{T} e^{-i\hat{H} t}\mathcal{T}^\dagger=e^{i\hat{H} t}$, as well as anti-commutes with all other operator generators of motion.  Its action on quantum states, $u,v$, is, $\mathcal{T}\vert u\rangle=\vert \bar u^*\rangle$, where the overbar represents the time reversed state and $*$ refers to the fact that the map is to the ``complex conjugate Hilbert space,"\cite{Schwinger} where $\langle u^*\vert v^*\rangle=\langle v\vert u\rangle$ and $\langle u^*\vert e^{-i\hat H t} \vert v^*\rangle=\langle v\vert e^{i\hat H t} \vert u\rangle$.
  
The vector potential has odd time reversal parity, so $\mathcal{T}\hat A_{\mu}(0,\mathbf{r})\mathcal{T}^\dagger=-\hat A_{\mu}(0,\mathbf{r})$.  Since $\hat A_{\mu}(t,\mathbf{r})=e^{i\hat{H}t}\hat{A}_\mu(0,\mathbf{r})e^{-i\hat{H}t}$, then $\mathcal{T}\hat A_{\mu}(t,\mathbf{r})\mathcal{T}^\dagger=-\hat A_{\mu}(-t,\mathbf{r})$.  It follows that:
\begin{align}
%\begin{split}
\langle g\vert \text{T}\big[&\hat{A}_\mu (t_2,\mathbf{r}_2) \hat{A}_\nu (t_1,\mathbf{r}_1)\big]\vert g\rangle \nonumber \\
&=\langle g\vert \mathcal{T}^\dagger\mathcal{T} \text{T} \big[\hat{A}_\mu(t_2,\mathbf{r}_2) \mathcal{T}^\dagger\mathcal{T} \hat{A}_\nu(t_1,\mathbf{r}_1)\big]\mathcal{T}^\dagger\mathcal{T}\vert g\rangle \nonumber \\
&=\langle \bar g^*\vert\text{T}\big[ \mathcal{T} \hat{A}_\mu(t_2,\mathbf{r}_2) \mathcal{T}^\dagger\mathcal{T} \hat{A}_\nu(t_1,\mathbf{r}_1)\mathcal{T}^\dagger\big] \vert \bar g^*\rangle  \nonumber\\
&=\langle \bar g^*\vert \text{T}\big[(-1)\hat{A}_\mu(-t_2,\mathbf{r}_2) (-1)\hat{A}_\nu(-t_1,\mathbf{r}_1)\big] \vert \bar g^*\rangle \nonumber \\
&=\langle \bar g\vert \text{T}\big[\hat{A}_\nu(-t_1,\mathbf{r}_1) \hat{A}_\mu(-t_2,\mathbf{r}_2)\big]\vert \bar g\rangle \nonumber\\
&=\langle \bar g\vert \text{T}\big[\hat{A}_\nu(t_2,\mathbf{r}_1) \hat{A}_\mu(t_1,\mathbf{r}_2)\big]\vert \bar g\rangle 
%\end{split}
\label{proof}
\end{align}
Where the last equality follows from time-translation symmetry.  It is then the case that if the ground state of the material is time reversal symmetric, $\vert \bar g\rangle=\vert g\rangle$, that
\begin{equation}
G^\text{F}_{\mu\nu}(t_2,\mathbf{r}_2,t_1,\mathbf{r}_1)=G^\text{F}_{\nu\mu}(t_2,\mathbf{r}_1,t_1,\mathbf{r}_2)
\label{result}
\end{equation}
There is a similar derivation of this symmetry for the retarded propagator, or else, it is obtained from analytic continuation of the above.

We refer to this result as the ``Reciprocity Theorem," and it is only satisfied when the ground state possesses microscopic time reversal symmetry.  Again, the restriction to point-like dipole sources is unnecessary, as an extended source is described by integrating $\mathbf{r}_1$ and $\mathbf{r}_2$ over the respective volumes.  Linear absorption in the sample is inconsequential; while the transition amplitude for absorption of a photon from $\mathbf{r}_1$ may be different for the amplitude of absorption for a photon from $\mathbf{r}_2$, these amplitudes are not measured and do not contribute to \ref{green}.  Finally, Onsager's relations\cite{Onsager1,Onsager2,Casimir,Kubo1,Kubo2,Efremov,Callen1,Callen2,Callen3,deGroot,Mazur} for linear response, the Rayleigh-Carson Electromagnetic Reciprocity Theorem,\cite{Rumsy, deHoop2, Carson,Lacomba} and its quantum counter-part for unitary evolution,\cite{LandauQuant,Weinberg,Coester,Coeste2} are all known manifestations of Equation \ref{result}.

Some scattering-matrix formulations of the theorem claim to satisfy reciprocity only in the asymptotic, far-field limit.\cite{deHoop,Saxon,Newton}  This is because scattering-matrix elements define transition amplitudes between free-space, plane-wave electromagnetic fields in the asymptotic past or future, and the perturbation expansion of the S-matrix is made in powers of the scattering material's contribution to the Hamiltonian.\cite{Fetter}  We approach reciprocity from a different perspective and find no such restriction, as we evaluate expectation values of the propagator with respect to $\vert g\rangle$ and perturbatively expand the measured quantity, $\mathbf{A}(t,\mathbf{r})$, in powers of the optical source's semi-classical contribution to the Hamiltonian, $\hat A_\mu(t,\mathbf{r}) \text{J}_\mu(t,\mathbf{r})$, as in linear response.\cite{Kubo1,Kubo2}  

To conclude the proof, circular polarization states may be represented by linear states via $\hat A_{\pm}(0,\mathbf{r})=\tfrac{1}{\sqrt{2}}\big(\hat A_{x}(0,\mathbf{r})\pm i\hat A_{y}(0,\mathbf{r})\big)$, so $\mathcal{T}\hat A_{+}(0,\mathbf{r})\mathcal{T}^\dagger=-\hat A_{-}(0,\mathbf{r})$.  This is sensible since the $\pm$ photon polarization states are eigenstates of spin angular momentum and the time reversal operator reverses its direction.  We return to considering retarded propagators, as they describe evolution of the system forward in time; when there is time reversal symmetry, analytic continuation of Equation \ref{result} gives:
\begin{equation}
G^\text{ret}_{++}(t_2,\mathbf{r}_2,t_1,\mathbf{r}_1)=G^\text{ret}_{--}(t_2,\mathbf{r}_1,t_1,\mathbf{r}_2)
\label{result2}
\end{equation}
Then there can not be Kerr rotation, as the frequency domain propagators are also the same, so $\theta_K=\tfrac{1}{2}\arg G^\text{ret}_{++}(\omega;\mathbf{r}_2,\mathbf{r}_1)-\tfrac{1}{2}\arg  G^\text{ret}_{--}(\omega;\mathbf{r}_1,\mathbf{r}_2)=0$.

This result does not always hold for nonlinear response, because the reflection amplitudes are not related by time reversal symmetry.  Consider a non-parametric process where the reflection of $+$ polarized light results in a spin excitation $\langle e_\uparrow\vert=\underset{t\rightarrow\infty}{\lim} \langle g\vert\hat e_\uparrow(t)$.  If the equilibrium state is time reverse symmetric, $\vert g\rangle=\vert \bar g\rangle$ and $\mathcal{T} \hat{e}_\downarrow(t)\mathcal{T}^\dagger=\hat{e}_\uparrow(-t)$, where $\vert e_\downarrow \rangle=\underset{t\rightarrow-\infty}{\lim} \hat {e}^\dagger_\downarrow(t)\vert g\rangle$, then applying $\mathcal{T}$ to a higher order propagator\cite{Efremov,Peterson,Callen1}  yields:
\begin{align}
\label{photocurrent}
\langle &e_\uparrow\vert \text{T}[\hat{A}_+(t_3,\mathbf{r}_2)\hat{A}_+ (t_2,\mathbf{r}_1)\hat{A}_+ (t_1,\mathbf{r}_1)]\vert g\rangle\\
&=-\langle g\vert \text{T}[\hat{A}_- (t_3,\mathbf{r}_1)\hat{A}_- (t_3+t_1-t_2,\mathbf{r}_1)\hat{A}_-(t_1,\mathbf{r}_2)]\vert e_\downarrow\rangle \nonumber
\end{align}
In other words, the amplitude of a process that results in the creation of an excited state of the material for light going from $\mathbf{r}_1\rightarrow\mathbf{r}_2$ is equal to that where an initial excited state decays and emits a photon for light going from $\mathbf{r}_2\rightarrow\mathbf{r}_1$.  Because the optical field is perturbing the material from the unilluminated equilibrium state in the infinite past, although the nonlinear excitation and decay processes are both possible, the Boltzmann weight for the material beginning in the excited state will be less than that of the ground state.  Although the two amplitudes above are equal up to a sign, their weightings are different and so there may be an asymmetry with respect to time reversal of the sum of all weighted amplitudes for light going from $\mathbf{r}_1\rightarrow\mathbf{r}_2$ and vice versa.  Kerr rotation may then be measured even if the equilibrium state of the material is time reverse symmetric.  This Kerr angle will be intensity dependent and if, as intensity is tuned to zero, the Kerr angle also approaches zero, then the equilibrium state of the material is necessarily time reverse symmetric.  Nonreciprocity is also possible if the spectral content of the incident and reflected beams differ, as when there is harmonic generation or Raman shifts.  This is demonstrated in a similar manner to that of the above; if light of frequency $\omega_1$ reflects to light of frequency $\omega_2$, there is no condition that the source for the incident light is such that the spectral weights for $\omega_1$ and $\omega_2$ are the same.  Thus, the two weighted sums of amplitudes will differ.

Non-equilibrium systems, such as a relaxing glass or a system driven by some other external source field, are inherently changing as a function of time, and so can give rise to nonreciprocity.  However, there is a subtlety in that $\hat H$ is the Hamiltonian for the whole world, so it is inaccurate, in this argument, to speak of open systems that break time-translation symmetry and invalidate the last step in \ref{proof}.  In other words, Equation \ref{result2} fails when microscopic time reversal symmetry is broken, but does not distinguish between systems in which it is broken due to a phase of matter that arises from spontaneous symmetry breaking, or from an external forcing, as in the Spin Hall Effect,\cite{kato1} where an applied current results in an unbalanced population of spins.  Likewise, there might be a highly excited state of a material that breaks mirror symmetry and emits radiation, as it relaxes, asymmetrically in the two circular polarization states, again, leading to an unbalanced spin population.  If these non-equilibrium systems are steady state,\cite{Coleman} then there will still be a density matrix, $\hat\rho$, that is not Boltzmann and is used to evaluate Equation \ref{green}.  Unless this density matrix manifestly breaks time reversal symmetry, $[\mathcal{T},\hat\rho]\neq 0$, then the measurement will satisfy reciprocity and there can be no Kerr rotation.

\section{Conclusion}
In proving Equation \ref{result2}, we have dispelled some incorrect ideas, recently promulgated, \cite{Pavan,Mineev,Arfi,Svirko1,Svirko2,Svirko3,Svirko4,Svirko5,Svirko6,Svirko7,Svirko8,Ginzberg0,Agranovich} as well as affirm and clarify the work of a number of studies.\cite{shelankov,dodge,armitage,Gridnev,Canright,Canright3,Cardona2,silverman,silverman2,silverman_expt,Halperin,Agronovich2,Schlaghneck,Vinogradov,Gridnev2,bokut,Fedorov,Nelson1,Nelson2,Cardona,Kamenetskii}  To summarize: (1) Kerr rotation may only arise from microscopic time reversal symmetry breaking; as will circular dichroism in normal incidence reflection.  This symmetry breaking may occur either through spontaneous symmetry breaking or by non-equilibrium processes.  Optically active materials, such as those with a $k$-linear susceptibility or any other form of mirror symmetry breaking, can not give rise to Kerr rotation, as they are time reversal symmetric.  (2) The proof above coincides with Onsager's relations and the Electromagnetic Reciprocity Theorem, and all three will fail only when microscopic time reversal symmetry is broken.  The theorems do not apply for nonlinear response, however nonlinear response must exhibit intensity dependent observables, such as Kerr rotation, or an alteration in the reflected frequency spectrum.  There are nonlinear effects that are intensity-independent and only alter the spectral content, such as spontaneous Raman shifts or spontaneous parametric photon down-conversion, but these effects are incoherent and yield a random phase delay.

These results constrain the predictions of \emph{all} constitutive relations\cite{lekner} used to model time reversal symmetric media.  A common source of confusion impeding the acceptance of these arguments have been calculations of Kerr rotation when using the mirror symmetry breaking, k-linear constitutive relations, with material constants allowed to vary with position:\cite{Svirko6,silverman2} $\mathbf{B}=\mathbf{H}$ and $\mathbf{D}=\epsilon^0(\mathbf{r})\mathbf{E}+\gamma(\mathbf{r})\nabla\times\mathbf{E}$ or $\mathbf{D}=\epsilon^0(\mathbf{r})\mathbf{E}+\nabla\times\left(\gamma(\mathbf{r})\right)\mathbf{E}$, where $\epsilon^0$ is the isotropic permittivity and $\gamma$ is the spatially dependent, isotropic gyrotropic parameter.\cite{LandauEM}  The resolution of this paradox is that only when the material constants are homogeneous will these relations conform to the intended symmetries of the model.  When there is a surface or spatial inhomogeneity, these constitutive relations do not satisfy Onsager's relations,\cite{Cardona,Cardona2} which means they do not explicitly satisfy time reversal symmetry and can not appropriately describe the system under discussion.  Furthermore, in lossless media, they don't respect Poynting's Theorem\cite{Fedorov,Lakhtakia} or follow from a least action principle.\cite{FriedThesis}  

Onsager's relations must be enforced if time reversal symmetric media are to be modeled correctly.  Consider the following permittivity tensor:
\begin{equation}
\begin{split}
\epsilon_{\mu\nu}(\omega,\mathbf{r},\mathbf{r}')=&\epsilon^0_{\mu\nu}(\omega,\mathbf{r})\delta(\mathbf{r}-\mathbf{r}')\\
&-\gamma_{\mu\nu\lambda}(\omega,\mathbf{r},\mathbf{r}')\partial_\lambda \delta(\mathbf{r}-\mathbf{r}')
\end{split}
%\label{gyro}
\end{equation}
This form generalizes the constitutive relations for k-linear response in homogeneous media, where $\gamma_{\mu\nu\lambda}$ and $\epsilon_{\mu\nu}^0$ will be constant, to a form where they are spatially dependent.  Onsager's relations, $\epsilon_{\mu\nu}(\omega,\mathbf{r},\mathbf{r}')=\epsilon_{\nu\mu}(\omega,\mathbf{r}',\mathbf{r})$, require that $\epsilon^0_{\mu\nu}(\omega,\mathbf{r})=\epsilon^0_{\nu\mu}(\omega,\mathbf{r})$ and $\gamma_{\mu\nu\lambda}(\omega,\mathbf{r},\mathbf{r}')=-\gamma_{\nu\mu\lambda}(\omega,\mathbf{r}',\mathbf{r})$.  For isotropic media, $\epsilon_{\mu\nu}^0(\omega,\mathbf{r})=\epsilon^0(\mathbf{r})$ and $\gamma_{\mu\nu\lambda}(\omega,\mathbf{r},\mathbf{r}')\equiv\varepsilon^{\mu\nu\lambda}\eta(\mathbf{r},\mathbf{r}')$, where $\eta(\mathbf{r},\mathbf{r}')$ is a scalar symmetric function.  As an example, if $\eta(\mathbf{r},\mathbf{r}')=\gamma(\tfrac{1}{2}\mathbf{r}+\tfrac{1}{2}\mathbf{r}')$, where $\gamma(\mathbf{r})$ is some other scalar function, then $\mathbf{D}=\epsilon^0(\mathbf{r})\mathbf{E}+\tfrac{1}{2}\gamma(\mathbf{r})\nabla\times\mathbf{E}+\tfrac{1}{2}\nabla\times(\gamma(\mathbf{r})\mathbf{E})$.  It can be easily checked that this form does not predict Kerr rotation,\cite{Agronovich2,Schlaghneck,Vinogradov,bokut,Fedorov} but our proof of Equation \ref{result2} guaranteed that this would be the case for \emph{any} choice of $\eta(\mathbf{r},\mathbf{r}')$ that is symmetric in the arguments, as Onsager's relations are correctly included.

The Sagnac Interferometer,\cite{spielman1,kapitulnik,dodge,Fried,xia1,lefevre} the instrument used to measure the Kerr angle in Karapetyan \emph{et.al.}, being a unique test for reciprocity, only measures microscopic time reversal symmetry breaking.  This is so because the interferometer measures the Kerr angle by interfering two beams of light made to reflect from the sample in a fashion such that the sourcing aperture for one beam is the receiving aperture for the other, and vice versa.  The Sagnac Interferometer conveys light of two linear polarization states to the sample by a polarization maintaining, single-mode optical fiber.  The end-face of the fiber is an ``aperture" for the two linear polarization states and the two modes that couple from free-space to the two fiber axes are the time reverse of those two that are emitted from it.  A quarter-wave plate, with slow axis oriented at $45^\circ$ with respect to the two polarization states emerging from the fiber axes, is placed between the fiber end-face and the sample.  The two orthogonal linearly polarized beams of light emitted from the fiber are transformed into opposite circularly polarized states after traversing the quarter-wave plate.  The circularly polarized beams of light partially reflect from the sample into the same circular polarization states and will pass through the quarter-wave plate a second time, transforming back into orthogonal linear polarization states, but now rotated $90^\circ$ from before.  In this way, the beams couple from one axis of the fiber to the other and interfere at a polarizer, oriented at $45^\circ$ with respect to both axes of the fiber, placed at the other end of the fiber optic cable.  A lock-in amplifier technique recovers the Kerr angle from the interference intensity.\cite{Fried}  Because the fiber is highly birefringent and the diode light source has $8\mu$m coherence length, only light that couples, after reflecting from the sample, between \emph{different} axes in the fiber will traverse optical path lengths that differ by less than a coherence length and interfere coherently at the polarizer.\cite{Fried}

The Reciprocity Theorem applies to the Sagnac interferometer exactly.  The spatial filtering of the fiber ensures that the electromagnetic spatial modes that are sourced and received by the fiber are exactly the time reverse of each other.  Comparing the phase delays of light exchanged between the two fiber axes uniquely tests for microscopic time reversal symmetry breaking, not only in a sample being probed, but also within the optical components that make up the instrument itself.  Misalignments or imperfect optical components will not introduce spurious signals, as they will have time reversal symmetric responses.

Because of the Reciprocity Theorem, the suggestion\cite{Pavan} that the recent measurements of a Kerr effect\cite{hovo} stem from an equilibrium phase of matter, with mirror symmetry breaking and without time reversal symmetry breaking, can not be correct.  Instead, the Reciprocity Theorem implies that either the ground state must break time reversal symmetry or the sample is in a highly non-equilibrium state that does as well.  More tests are needed to determine if nonlinear effects are relevant.

\section{Acknowledgments} This work would not have been possible without the support and insightful feedback of Robert Laughlin and Martin Fejer.  We also acknowledge helpful discussions with  Weejee Cho, Aharon Kapitulnik, Steven Kivelson, Srinivas Rhagu, Hovnatan Karapetyan, Pavan Hosur, Shanhui Fan, Samuel Lederer and Vladimir Mineev.  We are grateful for support from the Center for Probing the Nanoscale, NSF NSEC Grant PHY-0830228.

\bibliographystyle{unsrt}

\end{document}